\documentclass{bmcart}

\usepackage[utf8]{inputenc} 

\usepackage{pdflscape}
\usepackage{amsmath}
\usepackage{numprint} 
\usepackage{color, colortbl}
\usepackage{hyperref}
\usepackage[labelfont=bf,labelsep=period,justification=raggedright]{caption}
\usepackage[morefloats=7]{morefloats}

\def\includegraphics{}

\startlocaldefs
\endlocaldefs

\newcommand{\ERquantity}{Epi-feature}
\newcommand{\ERquantities}{Epi-features}
\newcommand{\quantities}{features}

\newcommand{\metrics}{measures}
\newcommand{\Metrics}{Measures}
\newcommand{\metric}{measure}

\newcommand	{\ERquantityb}{Epi-feature }
\newcommand{\ERquantitiesb}{Epi-features }
\newcommand{\quantitiesb}{features }
\newcommand{\quantityb}{feature }
\newcommand{\metricsb}{measures }
\newcommand{\Metricsb}{Measures }
\newcommand{\metricb}{measure }
\newcommand{\Metricb}{Measure }

\definecolor{purple1}{rgb}{0.45,0.12,.54}
\definecolor{Black}{rgb}{0,0,0}

\newcommand{\Revision}[1]{\textcolor{Black}{#1}}

\definecolor{R1}{rgb}{0.18,0.55,0.34} 
\definecolor{R2}{rgb}{0.4,0.80,0.67}  
\definecolor{R3}{rgb}{0.50,1 , 0.83} 
\definecolor{R4}{rgb}{1,0.89,0.77} 
\definecolor{R5}{rgb}{0.96,0.64,.38} 
\definecolor{R6}{rgb}{0.82,0.41,0.18} 

\begin{document}

\begin{frontmatter}

\begin{fmbox}
\dochead{Technical advance article}



\title{A Framework for Evaluating Epidemic Forecasts}

\author[
   addressref={aff1,aff3},                   
   corref={aff1},                       
   email={fstaba2@vt.edu}   
]{\inits{FST}\fnm{Farzaneh Sadat} \snm{Tabataba}}
\author[
   addressref={aff2},
   email={prithwi@vt.edu}
]{\inits{PC}\fnm{Prithwish } \snm{Chakraborty}}
\author[
   addressref={aff1,aff3},
   email={naren@cs.vt.edu  }
]{\inits{NR}\fnm{Naren} \snm{Ramakrishnan}}
\author[
   addressref={aff3},
   email={vsriniv@bi.vt.edu  }
]{\inits{SV}\fnm{Srinivasan } \snm{Venkatramanan}}
\author[
   addressref={aff3},
   email={chenj@bi.vt.edu  }
]{\inits{JC}\fnm{Jiangzhuo } \snm{Chen}}
\author[
   addressref={aff3},
   email={blewis@bi.vt.edu  }
]{\inits{BL}\fnm{Bryan} \snm{Lewis}}
\author[
   addressref={aff1,aff3},
   email={mmarathe@vt.edu  }
]{\inits{MM}\fnm{Madhav} \snm{Marathe}}

\address[id=aff1]{
  \orgname{Computer Science Department, Virginia Tech }, 
  \street{2202 Kraft Drive},                     %
  \postcode{24060}                                
  \city{Blacksburg/Virginia},                              
  \cny{USA}                                    
}
\address[id=aff2]{%
  \orgname{Computer Science Department, Virginia Tech},
  \street{7054 Haycock Road
 Falls Church},
  \postcode{22043}
  \city{Virginia},
  \cny{USA}
}
\address[id=aff3]{%
  \orgname{Network Dynamics and Simulation Science Laboratory (NDSSL),Biocomplexity Institute, Virginia Tech},
  \street{1015 Life Science Cir},
  \postcode{24061}
  \city{Blacksburg/Virginia},
  \cny{USA}
}


\begin{artnotes}
\end{artnotes}

\end{fmbox}


\begin{abstractbox}

\begin{abstract} 
\parttitle{Background} 
Over the past few decades, numerous forecasting methods have been proposed in 
the field of epidemic forecasting. Such methods can be classified into different
categories such as deterministic vs. probabilistic, comparative methods
vs. generative methods, and so on.
In some of the more popular comparative methods, researchers compare observed epidemiological data from
early stages of an outbreak with the output of proposed models to forecast the future trend
and prevalence of the pandemic.
A significant problem in this area is the lack of standard well-defined evaluation measures to select the best algorithm among different ones, as well as for selecting the best possible configuration for a particular algorithm.

\parttitle{Results} 
In this paper we present an evaluation framework which allows for combining different \quantities, error \metrics, and ranking schema to evaluate forecasts. We describe the various epidemic \quantitiesb (\ERquantities) included to characterize the output of forecasting methods and provide suitable error \metricsb that could be used to evaluate the accuracy of the methods with respect to these \ERquantities. We focus on long-term predictions rather than short-term forecasting and demonstrate the utility of the framework by evaluating six forecasting methods for predicting influenza in United States. Our results demonstrate that different error \metricsb lead to different rankings even for a single \ERquantity. Further, our experimental analyses show that no single method dominates the rest in predicting all \ERquantities, when evaluated across error \metrics. As an alternative, we provide various consensus ranking schema that summarize individual rankings, thus accounting for different error \metrics.  Since each \ERquantityb   presents a different aspect of the epidemic, multiple methods need to be combined to provide a comprehensive forecast. Thus we call for a more nuanced approach while evaluating epidemic forecasts and we believe that a comprehensive evaluation framework, as presented in this paper, will add value to the computational epidemiology community. 



%

\end{abstract}



\end{abstractbox}
%

\end{frontmatter}



\section*{ Background}
There is considerable interest in forecasting about future trends in 
diverse fields such as weather, economics and epidemiology\cite{Gen:1,Gen:2,Gen:3,Gen:4,Gen:5,Gen:6}. Epidemic forecasting, specifically, is of prime 
importance to epidemiologists and health-care providers, and many forecasting methods
have been proposed in this area\cite{Shaman:Review}. 
Typically, predictive models receive input in the form of a time-series 
of the epidemiological data from early stages of an outbreak
and are used to predict a few data points in the future and/or the remainder of the season.
However, assessing the performance of a forecasting algorithm is a big
challenge. Recently, several epidemic forecasting challenges have been 
organized by Centers for Disease Control and Prevention (CDC), National 
Institutes of Health (NIH), Department of Health and Human Services (HHS), 
National Oceanic and Atmospheric Administration (NOAA), and Defense Advanced 
Research Projects Agency (DARPA) to encourage different research groups to 
provide forecasting methods for disease outbreaks such as 
Flu \cite{CDC:Challenge}, Ebola \cite{Ebola:Challenge}, 
Dengue \cite{Dengue:Challenge1,Dengue:Challenge2} and  
Chikungunya \cite{Chikungunya:Challenge}. 
Fair evaluation and comparing the output of different forecasting methods has 
remained an open question. Three competitions named Makridakis Competitions 
(M-Competitions), were held in 1982, 1993, and 2000 and intended to evaluate and 
compare the performance and accuracy of different time-series forecasting 
methods \cite{Makridakis1993,Makridakis2000}. In their analysis, the accuracy of 
different methods is evaluated by calculating different error \metricsb on 
business and economic time-series which may be applicable to other disciplines. 
The target for prediction was economic time-series which have characteristically different behavior 
compared to those arising in the epidemiology. Though their analysis is generic 
enough, it does not consider properties of the time-series that are 
epidemiologically relevant. 
\Revision{Armstrong} \cite{armstrong2001} provides a thorough summary of the key 
principles that must be considered while evaluating such forecast methods. Our 
work expands upon their philosophy of objective evaluation, with specific focus 
on the domain of epidemiology.  To the best of our knowledge, 
at the time of writing this paper, there have been no formal studies on 
comparing the standard epidemiologically relevant \quantitiesb across appropriate 
error \metricsb for evaluating and comparing epidemic forecasting algorithms.

Nsoesie \etal \cite{Elain:Review} reviewed different studies in the field
of forecasting influenza outbreaks and presented the \quantitiesb used to
evaluate the performance of proposed methods. Eleven of the sixteen forecasting methods studied by the authors, predicted daily/weekly case counts \cite{Elain:Review}. Some of the studies used various distance functions or errors as a measure of closeness between the predicted and observed time-series. For example, Viboud \etal \cite{E:10}, Aguirre and Gonzalez \cite{E-9}, and Jiang \etal \cite{E-11} used correlation coefficients to calculate the accuracy of daily or weekly forecasts of influenza case counts.
Other studies evaluated the precision and "closeness" of predicted activities to
observed values using different statistical measure of errors such as 
root-mean-square-error (RMSE), percentage error \cite{E-11,E-12}, etc. However, defining a good distance function which demonstrates closeness 
between the surveillance and predicted epidemic curves is still a challenge. Moreover, the distance function provides a general comparison between the two time-series and ignores their epidemiological relevance \Revision{ between them, i.e. specific features of the epidemic curves that are more significant and meaningful from the epidemiologist perspective; these features could be better criteria to compare epidemic curves together rather than simple distance error. }
Cha\cite{R-Distance} provided a survey on different distance/similarity functions for calculating the closeness between two time-series or discrete probability density functions.  Some other studies have analyzed the overlap or difference between the predicted and observed weekly activities by graphical inspection \cite{E-15}.
Epidemic peak is one of the most important quantities of interest in an outbreak, and its magnitude and timing are important from the perspective of health service providers. Consequently, accurately predicting the peak has been the goal of some forecasting studies \cite{E-2,E-3,E-4,E-5,E-9,E-15,E-16,E-18,E-20,E-21}.
Hall \etal \cite{E-3}, Aguirre and Gonzalez \cite{E-9} 
and Hyder \etal \cite{E-21} predicted the pandemic duration and
computed the error between the predicted and real value. 
A few studies also consider the attack rate for the epidemic season as the \quantityb of interest for their method \cite{E-12,E-5}. 
\subsection*{Study Objective \& Summary of Results}
In this paper, an epidemic forecast generated by a model/data-driven approach is to be quantified based on epidemiologically relevant \quantitiesb which we hereon refer to as \emph{\ERquantities}. Further, the accuracy of a model's estimate of a particular \ERquantitiesb is to be quantified by evaluating its error with respect to the \ERquantitiesb extracted from the ground truth. This is enabled by using functions that capture their dissimilarity, which we hereon refer to as \emph{error \metrics}.

We present a simple end to end framework for evaluating epidemic forecasts, keeping in mind the variety of epidemic \quantitiesb and error \metricsb that can be used to quantify their performance. The software framework, Epi-Evaluator (shown in Figure~\ref{fig:framework}), is built by taking into account several possible use cases and expected to be a growing lightweight library of loosely coupled scripts. To demonstrate its potential and flexibility, we use the framework on a collection of six different methods used to predict influenza in the United States. In addition to quantifying the performance of each method, we also show how the framework allows for comparison among the methods by ranking them. 

We used influenza surveillance data, as reported by United States Centers for Disease Control and Prevention (CDC)\cite{site-CDC}, as the gold standard epidemiological data. Output of six forecasting methods was used as the predicted data. We calculated 8 \ERquantitiesb on the 2013-2014 season data against 10 HHS regions of the United States (provided by U.S. Department of Health \& Human Services) \cite{HHS-regions} and 6  error \metricsb to assess the \ERquantities. We applied the proposed \ERquantitiesb and error \metricsb on both real and predicted data to compare them to each other.

As expected, the performance of a particular method depends on the \ERquantitiesb and error \metricsb of choice. Our experimental results demonstrate that some algorithms perform well with regards to one \ERquantityb but do not perform well with respect to other \quantities.  It is possible that none of the forecasting algorithms could dominate all the other algorithms in every \ERquantityb and error \metric. 

As one \Revision{\ERquantityb} cannot describe all \Revision{attributes} of a forecasting algorithm\Revision{'s output}, all of them should be considered in the ranking process to have a comprehensive comparison. We suggest aggregation of different error \metricsb in the ranking procedure. To this effect, we show how Consensus Ranking could be used to provide comprehensive evaluation. In addition, depending on the purpose of the forecasting algorithm, some \ERquantitiesb could be considered more significant than others, and weighted more accordingly while evaluating forecasts. We recommend  \Revision{a second level} of Consensus Ranking to accumulate the analysis for various \ERquantitiesb and provide a total summary of forecasting methods' capabilities.  
\\We also propose another ranking method, named Horizon ranking, to provide a comparative evaluation of the methods performance across time. If the Horizon Ranking fluctuates a lot over the time steps, that gives lower credit to the average Consensus Ranking as selection criteria for the best method. Based on experimental results of Horizon ranking, you will notice that for a single \ERquantity, one method may show the best performance in early stages of the prediction, whereas another algorithm is the dominator in other time intervals. Finding a pattern in Horizon Ranking plots helps to figure out which methods should be selected for different periods of forecasting.

Note that many of the proposed \ERquantitiesb or error \metricsb have been studied  
earlier in the literature. The aim of our study is to perform an objective 
comparison across \ERquantitiesb and error \metricsb and ascertain their impact on 
evaluating and ranking competing models. Further, the focus is not on the 
performance of methods being compared, but on the features provided by the 
software framework for evaluating them. The 
software package is scheduled to be released in an open source environment. We 
envision it as a growing ecosystem, where end-users, domain experts and 
statisticians alike, can contribute \ERquantitiesb and error \metricsb for 
performance analysis of forecasting methods. 


\section*{Methods}
The goal of this paper is demonstrating how to apply the \ERquantitiesb and 
error \metricsb on the output of a forecasting algorithm to evaluate its 
performance and compare it with other methods. We implemented \Revision{a stochastic compartment SEIR  
algorithm \cite{Lekone:2006}} with six different configurations to forecast influenza outbreak 
(described in the Supplementary material). These six configurations result in 
different forecasts which are then used for evaluation. In the following 
sections, we expand upon the different possibilities we consider for each module 
(\ERquantities, error \metricsb and ranking schema) and demonstrate their effect 
on evaluating and ranking the forecasting methods. 

\section*{Forecasting Process}
Epidemic data are in the form of a time-series such as
$y\left(1\right),..., y\left(t\right),..,y\left(\Revision{T}\right)$, where
$y\left(t\right)$ denotes the number of new infected cases observed in time
$t$, and \Revision{$T$} is the duration of the epidemic season. Weekly time-steps are usually preferred to average out the noise in daily case counts. 
Let denote the \textit{prediction time} by $k$ and the prediction horizon by \Revision{$w$}. Given the early time-series up to time $k$ ($y\left(1\right),..., y\left(k\right)$) as observed data, the forecasting algorithm predicts
the time-series up to the prediction horizon as $x\left(k+1\right),...,x \left(k+\Revision{w}\right)$. The forecasts could be short-term (small \Revision{$w$}), or long-term ($\Revision{w=T}-k$). As most of the proposed \ERquantitiesb are only defined based on the complete epidemic curve rather than a few predicted data points, we generate long-term forecasts for each prediction time.  The remainder of the observed time-series ($y\left(k+1\right),...,y \left(\Revision{T}\right)$) is used as a test set for comparing with the predicted time-series (Figure~\ref{fig:fig0}). We increment the prediction time $k$, and update the predictions as we observe newer data points. For each prediction time $k$, we generate an epidemic curve for the remainder of the season. 

\section*{Epidemiologically Relevant Features}
 In this section, we list the \ERquantitiesb we will use to 
characterize the features of an epidemic time-series.
While some of these quantities are generic and applicable to any time-series, the others are specific to epidemiology. Table~\ref{tb:notations} summarizes the notations needed to define these \ERquantities. 

\begin{table*}[h!]
\centering
\caption{\label{tb:notations}  Notation and Symbols }
\begin{tabular}{cp{10cm}}
\hline 
\textbf{Symbol} &  \ \ \ \ \ \ \textbf{Definition}
 \\
\hline
$y_t$  & number of new cases of disease in the $t^{th} $ week  observed in surveillance data  \\
\hline
$x_t$  & number of new cases of disease in the $t^{th} $ week predicted by forecasting methods   \\
$x_{start}$  & number of new cases of disease predicted at the start of epidemic season   \\
$x_{peak}$  & number of new cases of disease predicted at the peak of epidemic season    \\
\hline
$e_t $  & $e_t = y_t-x_t $ : the prediction error  \\
\hline
$T $  & duration of the epidemic season  \\
\hline
$ \bar{y} $    & $\bar{y}=\frac {1}{T} \sum_{t=1}^T (y_t) $ : the mean for \textit{y} values over $T$ weeks 
\\
\hline
$\sigma^2$ & $ \sigma^2=\frac {1}{T-1} \sum_{t=1}^T (y_t-\bar{y})^2 $ : The variance of \textit{y} values over $T$ weeks \\
\hline
$n_{t} $ & Total number of infected persons during specified period  \\ 
\hline 
$n_{ps} $ & The population size at the start of specified period\\
\hline\ 
$n_{t} \left(age\right) $ & Total number of infected persons with specific age during the specified period  \\
\hline
$n_{ps} \left(age\right) $ & The population size with specific age at the start of specified period \\
\hline
$n_{c} $ & or $ n_{contacts} $ is the number of contacts of primary infected persons\\
\hline
$n_{sg}  $ &  or  $ n_{second-generation} $ is the new number of infected persons among the contacts of primary infected individuals during a specified period \\
\hline
$GM\{Error\} $ & $ GM(e)= (\prod_{i=1}^n(e_i))^{(1/n)} $ : Geometric Mean of a set of Errors \\
\hline
$M\{Error\} $ &  Arithmetic Mean of a set of Errors\\
\hline
$Md\{Error\} $ &  Median value a set of Errors\\
\hline
$RMS\{Error\} $ &  Root Mean Square of a set of Errors\\
\hline
\end{tabular}
\end{table*}

\subsection*{Peak Value \& Time}
Peak value is the highest value in a time-series. In the epidemic context, it
refers to the most number of newly infected individuals on any given week during an epidemic season. Closely associated with peak value, is peak time, which is the week in which the peak value is attained. Predicting these values accurately will help the healthcare providers estimate the resource burden and the preparation time. 

\subsection*{First-Take-Off (Value \& Time):}
Seasonal outbreaks, like the flu, usually remain dormant and exhibit a sharp rise in the number of cases just as the season commences. A similar phenomenon of sharp increase is exhibited by emerging infectious diseases. Referred to as "first-take-off" time, this is useful to detect early and will help the authorities alert the public and raise awareness.  Mathematically, it is the time at which the first derivative of the epidemic curve exceeds a specific
threshold. Since the epidemic curve is discretized in weekly increments, the approximate slope of the curve over $\Delta t$ time steps is defined as follows:

\begin{eqnarray}
s(x,\Delta t)	= \frac{x(t+\Delta t)-x(t)}{\Delta t}
\end{eqnarray}
where $x$  is the number of new infected case-counts and $t$ indicates the week number. In our experiment, we set $\Delta t=2$. The value of $s(x,\Delta t)$ is the slope of the curve and shows the take-off-value while the start time of the take-off indicates the take-off-time. \Revision{The threshold used in calculating the first-take-off depends on the type of the disease and how aggressive and dangerous the outbreak could be. The epidemiologists determine the threshold value. In this case, we set the threshold to 150. }

\begin{table*}[tb]
\centering
\caption{\label{tb:quantities}  Definitions of different Epidemiologically Relevant \quantities }
\begin{tabular}{cp{9cm}}
\hline 
\textbf{\ERquantityb name} &  \ \ \ \ \ \ \textbf{Definition}
 \\
\hline
Peak value  & Most number of weekly newly infected cases in the epidemic time-series  \\
\hline
Peak time & The week when peak value is attained \\
\hline
Total attack rate & Fraction of individuals ever infected in the whole population\\
\hline
Age-specific attack rate   & Fraction of individuals ever infected belonging to a specific age window
\\
\hline
First-take-off-(value): & Sharp increase in the number of new infected case counts over a few consecutive weeks  \\
\hline
First-take-off-(time): & The start time of sudden increase in the number of new infected case counts  \\ 
\hline 
Intensity duration & The number of weeks (usually consecutive) where the number of  new infected case counts is more than a specific threshold\\
\hline\ 
Speed of epidemic & Rate at which the case counts approach the peak value\\
\hline
Start-time of disease season & Time at which fraction of infected individuals exceeds a specific threshold \\
\hline
\end{tabular}
\end{table*}

\subsection*{Intensity Duration}
Intensity Duration (ID) indicates the number of weeks, usually consecutive,
where the number of new infected case counts is more than a specific threshold.
This \quantityb can be used by hospitals to estimate the number of weeks for which the epidemic will stress their resources (Figure \ref{fig:fig1}).

\subsection*{Speed of Epidemic}
The Speed of Epidemic (SpE) indicates how fast the infected case counts reach the peak value. This \quantityb includes  peak value and peak time simultaneously.
The following equation shows the definition of speed of epidemic: 

\begin{equation}
	SpE = \frac{x_{peak} - x_{start}}{t_{peak} - t_{start}}
\end{equation}
where $x_{peak}$ and $x_{start}$ are the number of new case count diseases at
peak time and the start time of the season, respectively.  In other words,
speed of epidemic is the steepness of the line that connects the start
data-point of time-series sequence to the peak data-point(Figure \ref{fig:fig2}).

\subsection*{Total Attack Rate (TAR):}
Attack rate (TAR) is the ratio of the total number of infected cases during a specified period, usually one season, to the size of the whole population at the start of the period. 
\begin{eqnarray}
\label{eq:}
\centering
TAR= \frac{n_{t} }{n_{ps}}
\end{eqnarray}
where $n_{t}$ is the total number of infected people during specified
period.

\subsection*{Age-specific Attack Rate (Age-AR)}
This is similar to the total attack rate but focuses on a specific
sub-population. Specific attack rate is not only limited to age-specific attack rate, but the sub-population could be restricted by any feature like age, gender, or any special group.
\begin{eqnarray}
	AgeAR\left(age\right) = \frac{n_{t}\left(age\right) }{n_{ps}\left(age\right)}
\end{eqnarray}

\subsection*{Secondary Attack Rate (SAR):} 
Secondary attack rate (SAR) means the ratio of new infected cases of a disease,
during a particular period, among the contacts of primary cases who are infected
first; In other words, it is a measure of the spreading of disease in the
contact network.
\begin{eqnarray}
	SAR = \frac{n_{sg} }{n_{c}}
\end{eqnarray}
where $n_{c}$ is the number of contacts of primary infected persons and
$n_{sg}$ is the number of infected persons among those contacts
during a specified period\cite{secondAR}.  In order to calculate the secondary
attack rate, individual information about households and their contacts network are needed. Epidemiologists estimated the secondary attack rate in
household contacts of several states of the United States which was $18\%$ to $19\%$ for acute-respiratory-illness (ARI) and $8\%$ to $12\%$ for influenza-like-illness
(ILI)\cite{CDC-AR}. 

\subsection*{Start-time of a disease Season}
\Revision{ We define the "Start-time of a flu season" as the week when the flu-percentage exceeds a specified threshold. The flu-percentage is defined as follows:}
\begin{eqnarray}
	Per\left(Flu\right) = \frac{n_{i}\left(Flu\right) }{n_{i}\left(All\right)}
\end{eqnarray}
\Revision{where $n_{i} \left(Flu\right) $ is weekly influenza related illnesses in
$i^{th}$ week and $n_{i}\left(All\right)$ is the weekly number of non-ILI patients seen by health providers for any reason and/or specimens recognized as negative cases by clinical laboratories. } 
Usually, predicting the denominator of
the above equation is a difficult task. But if the observed data is available, one
can use that to calculate the denominator
and only predicts the numerator of the above equation and calculates the
Flu percentage. \Revision{ The value of threshold that is used as the criteria is determined by epidemiologist and could be calculated in different ways. We define the threshold through the analysis of past flu seasons inspiring from the flu baseline definition by CDC \cite{cdc:baseline}. 
CDC defines the baseline as the mean percentage of visits for influenza during non-influenza weeks for the previous three seasons plus two standard deviations \cite{cdc:baseline}. 
The non-influenza weeks are defined as two or more consecutive weeks in which the number of counted ILI diagnoses for each week is less than 2\% of total seasonal ILI case counts. 
The definition of start-of-season could be generalized for any disease like Ebola, Zika, etc. }

\section*{\label{sec:metricSection}Error \Metrics}
 The second step of evaluating epidemic forecasting algorithms is to measure the error for each predicted \ERquantities. There are a variety of \metricsb that can be used to assess the error between the predicted time-series and the observed one. 
The error \metricsb that we consider in this study are listed in Table~\ref{tb:metrics} along with their features. The notations used in the error \metricsb equations are described in Table~\ref{tb:notations}. Note that all the error \metricsb considered, only handle the absolute value of the error. They do not distinguish between under and over-estimation of the time-series.  The signed versions of some of these absolute error measures are listed in the supporting information. These signed \metricsb include the direction of error i.e. the positive sign demonstrates the underestimation while the negative one indicates
overestimation. Moreover, all the \metricsb referred to in Table~\ref{tb:metrics} use Arithmetic Mean to get an average value of the error. Variants that use geometric mean, median, etc. are listed in the supporting information section. 


\begin{landscape}
\begin{table*}[h!]
\centering
\caption{\label{tb:metrics}  List of main Error \Metricsb. Arithmetic mean and absolute errors are used to calculate these \metricsb in which positive and negative deviations do not
cancel each other out and \metricsb do not provide any information about the direction of errors. }
\begin{tabular}{p{2.8cm}|c|p{3.5cm}cp{1cm}p{1.5cm}p{1.2cm}p{3cm}}
\hline 
\Metricb name & Formula & Description & Scaled & Outlier Protection & Other forms & Penalize extreme deviation & Other Specification\\
\hline
Mean Absolute Error (MAE) & $$$ MAE=\frac {1}{T} \sum_{t=1}^T |e_t| $$$ & Demonstrates the magnitude of overall error & No & Not Good & GMAE & NO &  - \\
\hline 
Root Mean Squared Error (RMSE) & $$$ RMSE= \sqrt{ \frac {\sum_{t=1}^T e_t^2}{T} } $$$ & Root square of average squared error & No & Not Good & MSE & Yes & - \\
\hline
Mean Absolute Percentage Error (MAPE) & $$$ MAPE=\frac {1}{T} \sum_{t=1}^T |\frac{e_t}{y_t}|  $$$ & Measures the percentage of average absolute error & Yes & Not Good & MdAPE\footnote{Md represent Median}, RMSPE\footnote{RMS represent Root Mean Square}  & NO &  - \\
\hline
symmetric Mean Absolute Percentage Error (sMAPE) & $$$ sMAPE=\frac {2}{T} \sum_{t=1}^T |\frac{e_t}{y_t+x_t}| $$$ & Scale the error by dividing it by the average of $y_t$ and $x_t$   & Yes & Good & sMdAPE & No & Less possibility of division by zero rather than MAPE.  \\

\hline
Mean Absolute Relative Error (MARE) & $$$ MARE=\frac {1}{T} \sum_{t=1}^T |\frac{e_t}{e_{RWt}}|  $$$ & Measures the average ratio of absolute error to Random walk error & Yes & Fair & MdRAE, GMRAE & No & - \\

\hline
Relative Measures: e.g. RelMAE (RMAE) & $$$ RMAE=\frac {MAE}{MAE_{RW}}= \frac{\sum_{t=1}^T |e_t|}{\sum_{t=1}^T |e_{RWt|} }|  $$$ & Ratio of accumulation of errors to cumulative error of Random Walk method & Yes & Not Good & RelRMSE, LMR \cite{Survey2013} , RGRMSE \cite{Survey17}& No & - \\

\hline
Mean Absolute Scaled Error (MASE) & $$$ MASE=\frac {1}{T} \sum_{t=1}^T |\frac{e_t}{\frac{1}{T-1}\times\sum_{i=2}^T|y_i-y_{i-1}|}|  $$$ & Measures the average ratio of error to average error of one-step Random Walk method  & Yes & Fair & RMSSE  & No & - \\

\hline
Percent Better (PB)    & $$$ PB=\frac {1}{T} \sum_{t=1}^T [I\{e_t,e_{WRt}\}] $$$  & Demonstrates average number of times that method overcomes the Random Walk method & Yes & Good & - & No & Not good for calibration and close competitive methods.   \\
& $$$ |e_{s,t}|\leq|e_{WRt}| \leftrightarrow I\{e_{t},e_{WRt}\}=1 $$$ & & & & & & \\

\hline
Mean Arctangent Absolute Percentage Error (MAAPE) & $$$ MAAPE=\frac {1}{T} \sum_{t=1}^T arctan|\frac{e_t}{y_t}|  $$$ & Calculates the average arctangent of absolute percentage error & Yes & Good & MdAAPE & No & Smooths large errors. Solve division by zero problem.  \\
%
\hline\ 
Normalized Mean Squared Error (NMSE) & {$$$NMSE=\frac {MSE}{\sigma^2} = \frac {1}{\sigma^2 T} \sum_{t=1}^T e_t^2 $$$} & Normalized version of MSE: value of error is balanced & No & Not Good & NA & No & Balanced error by dividing by variance of real data. \\
\hline

\end{tabular}

\end{table*}
\end{landscape}

After careful consideration, we have selected MAE, RMSE, MAPE, sMAPE, MdAPE and MdsAPE as the error \metricsb for evaluating the \ERquantitiesb and ignored others based on different reasons. We list some of \Revision{ these reasons and observations on the eliminated error \metricsb in part B of Supplementary Information. } 
Also, instead of using MAPE, we suggest corrected MAPE (cMAPE) to solve 
the problem of division by zero:
 
\begin{eqnarray}
    cMAPE= 
\begin{cases}
    \frac {1}{T} \sum_{t=1}^T |\frac{e_t}{y_t}|,& \text{if } y_t\neq0\\
    \frac {1}{T} \sum_{t=1}^T |\frac{e_t}{y_t+\epsilon}|,              & \text{otherwise}
\end{cases}
\end{eqnarray}
where $\epsilon$ is a small value. It could be equal to the lowest non-zero 
value of observed data. We have also added two error \metricsb based on the 
median namely,  Median Absolute Percentage Error (MdAPE) and Median symmetric 
Absolute Percentage Error (MdsAPE). However, as median errors have low 
sensitivity to change in methods, we do not recommend them for isolated use as 
the selection or calibration criteria. 

\section*{\label{sec:ranking}Ranking Methods}
The third step of the evaluation process is 
ranking different methods based on different \ERquantitiesb and the result of 
different error \metrics. For this purpose, we have used two kinds of ranking 
methods: Consensus Ranking and Horizon Ranking. 

\begin{itemize}
\item \textbf{Consensus Ranking}: Consensus Ranking (CR) for each method is 
defined as the average ranking of the method among others. This kind of 
Consensus Ranking could be defined in different scenarios. For example, the 
average ranking that is used in Table~\ref{tb:peak_ranks} in Result section is 
Consensus Ranking of a method based on one specific \ERquantityb across 
different error \metrics.  
    
\begin{eqnarray}
    CR_{EM}^m=  \frac {1}{n_{EM}} \sum_{i=1}^{n_{EM}} |\frac{R_{i,m}}{n_{EM}}|
\end{eqnarray}

where $R_{i,m}$ is the individual ranking assigned to method $m$ among other 
methods for predicting one \ERquantityb  based on error \metricb $i$, and 
Consensus Ranking $CR_{EM}^m$ is the overall Ranking of method m based on 
different error \metrics. 

Consensus Ranking could also be defined across different \ERquantities. In this 
case, CR over error \metricsb could be considered as the individual 
ranking of a method, and the average is calculated over different 
\ERquantities. It is important to consider the variance of ranking and the 
intensity of quartiles besides the mean value of CR. In the Results section we 
demonstrate how to process and analyze these rankings in a meaningful way.

\item \textbf{Horizon Ranking}: While Consensus ranking considers the average
performance of methods over prediction times, Horizon ranking 
demonstrates the trend of forecasting methods' performance in predicting a 
single  \ERquantityb across different prediction times. First of all, for each \ERquantity, 
we compute an error \metricb like Absolute Percentage Error (APE) or its 
symmetric variant (sAPE) per prediction time. For each prediction time, APE 
values of different forecasting methods are sorted from smallest to largest to 
determine the ranking of the methods. The average value of this ranking over 
different error \metricsb determines the overall Horizon Ranking of the methods in 
each time-step. 
\end{itemize}

\section*{Data}
The ILI surveillance data used in this paper were obtained from the website of the United States Centers for Disease Control and Prevention (CDC). The information of patient
visits to health care providers and hospitals for ILI was
collected through the US Outpatient Influenza-like Illness Surveillance
Network since 1997 and lagged by two weeks(ILINet)\cite{26r,site-CDC}; 
This Network covers all 50 states, Puerto Rico, the District of Columbia and the U.S. Virgin Islands.
The weekly data are separately provided for 10 regions of
HHS regions \cite{HHS-regions} that cover all of the US. The forecasting algorithms have been applied to CDC data for each HSS region.
We applied our forecasting algorithm on the 2013-2014 flu season data where every season is less than or equal to one year and contains one major epidemic.
Figure~\ref{fig:Map} shows the HHS Region Map that assigned US states to the
regions. 

\section*{Results and Analysis}\label{sec:evalSection}
Past literature in the area of forecasting performs an overall evaluation for 
evaluating the performance of the predictive algorithm by defining a 
statistical distance/similarity function to measure the closeness of predicted 
epidemic curve to the observed epidemic curve. However, they rarely evaluate 
the robustness of a method's performance across epidemic features of interest 
and error \metrics. Although the focus of the paper is not on the method to be chosen, it is instructive to observe the software framework in action as we use different evaluation criterion for the methods.



\subsection*{Rankings based on Error \Metricsb applied to peak value}
In Table~\ref{tb:all-peak}, we have calculated six error \metrics, MAE, RMSE, 
MAPE, sMAPE, MdAPE, and MdsAPE for the peak value predicted by six different 
forecasting methods. The corresponding ranks are provided in the Ranking Table 
(Table~\ref{tb:peak_ranks}). The most successful method is assigned rank 1 
(R1); As can be seen, even similar \metricsb like MAPE and sMAPE do not behave 
the same for the ranking process. The fourth algorithm wins six first 
places among other methods for seven error \metricsb that shows almost the best 
performance. However, it is hard to come to a similar conclusion for other 
methods. The last column in the table is Consensus Ranking, which shows the 
average ranking of the method over different error \metrics. According to Consensus Ranking, some methods like method 2 and 5 could have close or same average ranking which makes the comparison harder.  
Figure~\ref{fig:peakBoxplot} 
shows the Box-Whisker diagram of methods' rankings. The second and third 
quartile area around the median of the ranking are much more intense for the fifth 
method than the second one which shows more certainty about the median value 
(4) as correct ranking. Also, the two data points which deviate the mean value 
from the median are recognized as outliers. However, for the second method, 
quartiles' boxes around the median are wider and whiskers are longer which 
implies less certainty about the mean and median value of the ranking. 
Based on such analysis, the fourth method (M4) is the superior for predicting 
the peak value. After that the order of performance for other methods will 
be: Method 6 (M6), Method 3, Method 5, Method 2 and Method 1. Note, however, 
that this analysis is specific to using peak value as the \ERquantityb of 
interest.

\begin{table*}[tb]
\centering
\captionsetup{font=scriptsize}
\caption{\label{tb:all-peak} Different errors for predicting peak value for Region 1 over whole season (2013-2014).}
\begin{tabular}{ccccccc}
\hline
         & {MAE } & {RMSE} & {MAPE }  & {sMAPE } & {MdAPE} & {MdsAPE} \\
\hline
Method 1 &  4992.0 & 9838.6 & 4.9  & 1.04 & 1.7 &  1.03  \\
\hline
Method 2 & 4825.2 & 9770.4 & 4.7  & 0.99 & 1.4  &  0.95 \\
\hline
Method 3 & 3263.0 & 5146.5 & 3.2  & 0.96 & 1.5 &  1.01 \\
\hline
Method 4 & 2990.7 &  4651.3 & 2.9  & 0.899 & 1.1 &  0.85 \\
\hline
Method 5 & 3523.2 & 5334.8 & 3.4  & 0.95 & 2.1  & 1.01 \\
\hline
Method 6 & 3310.9 & 4948.5 & 3.2  & 0.896 & 1.5 & 0.85 \\

\hline
\end{tabular}

\end{table*}


\begin{table*}[tb]
\centering
\captionsetup{font=scriptsize}
\caption{\label{tb:peak_ranks} Ranking of methods for predicting peak value based on different error \metricsb for Region 1 over whole season (2013-2014). \Revision{The color spectrum demonstrates different ranking levels. Dark green represents the best rank, whereas dark orange represents the worst one.}}
\begin{tabular}{cccccccp{1.2cm}c}
\hline
         & {MAE } & {RMSE} & {MAPE }  & {sMAPE } & {MdAPE} & {MdsAPE} & Consensus Ranking & Median \\
\hline
Method 1 &  \cellcolor{R6} 6 & \cellcolor{R6} 6 & \cellcolor{R6} 6  & \cellcolor{R6} 6 & \cellcolor{R5} 5 & \cellcolor{R6}  6 & \cellcolor{R6} 5.83 &\cellcolor{R6} 6\\
\hline
Method 2 & \cellcolor{R5} 5 & \cellcolor{R5} 5 & \cellcolor{R5} 5  & \cellcolor{R5} 5 & \cellcolor{R2}2  & \cellcolor{R3}  3  & \cellcolor{R4} 4.17 & \cellcolor{R5} 5\\
\hline
Method 3 & \cellcolor{R2} 2 & \cellcolor{R3} 3 & \cellcolor{R2}2  &  \cellcolor{R4} 4 & \cellcolor{R3} 3 & \cellcolor{R4} 4 & \cellcolor{R3} 3.00  & \cellcolor{R3} 3\\
\hline
Method 4 & \cellcolor{R1} 1 & \cellcolor{R1}1 &  \cellcolor{R1}1  & \cellcolor{R2} 2 & \cellcolor{R1}1 & \cellcolor{R1} 1 & \cellcolor{R1} 1.17 & \cellcolor{R1} 1 \\
\hline
Method 5 & \cellcolor{R4} 4 & \cellcolor{R4} 4 & \cellcolor{R4} 4  & \cellcolor{R3} 3 & \cellcolor{R6} 6  & \cellcolor{R4} 4 & \cellcolor{R4} 4.17 &\cellcolor{R4} 4 \\
\hline
Method 6 & \cellcolor{R3} 3 & \cellcolor{R2} 2 & \cellcolor{R3} 3  & \cellcolor{R1}1 &  \cellcolor{R3}3 &  \cellcolor{R1}1 & \cellcolor{R2} 2.17 &\cellcolor{R2} 2.5\\

\hline
\end{tabular}

\end{table*}

\subsection*{Consensus Ranking across all \ERquantities}
In order to make a comprehensive comparison, we have calculated the error 
\metricsb on the following \ERquantities: Peak value and time, Take-off-value and 
Take-off-time, Intensity Duration's length and start time, Speed of epidemic, 
and start of flu season. We do not include demographic-specific \ERquantitiesb 
such as age-specific attack rate or secondary attack rate, since such 
information \Revision{is not available for our methods}. 

Figure~\ref{fig:AllEpiBoxPlot} shows the 
Consensus Ranking of the methods in predicting different \ERquantitiesb for Region 1. Note that
Method 4, which is superior in predicting some \ERquantitiesb such as Peak value 
and start of Flu season, is worse than other methods in predicting other 
\ERquantitiesb such as  Take-off time and Intensity Duration. The tables 
corresponding to the box-plots are included in supporting information. \\ 
Figure~\ref{fig:CS_BoxPlot} shows the second level of Consensus Ranking over various \ERquantitiesb for Region 1. This figure   summarizes the performance of different methods based on the average consensus rankings that 
are listed in Table~\ref{tb:CS_all}. It is evident that the first, second and 
fifth methods have similar performance, while the third method performs 
moderately well across \ERquantities. The fourth method which performs best for 
five out of eight \ERquantities, however, is not among the top three methods for 
predicting Take-off time and Intensity Duration. Method 6 comes in as the 
second best method when considering the consensus ranking.
\Revision{
The first level of Consensus Ranking over error \metricsb for other HHS regions are included in Figures S2-Fig to S10-Fig of Supporting Information. Figures~\ref{fig:CS-ALL-R1-6} and~\ref{fig:CS-ALL-R7-10} represent the second level of Consensus Rankings of the six approaches over all \ERquantitiesb for regions 1 to 10. If each region can be assigned with a different method as predictor comparing other regions, we suggest to assess the performance of methods on each region separately and to select the best one. Otherwise, if the experts need to select one method as the best predictor for all regions, we propose the third level of Consensus Ranking to aggregate the results across different regions. Figure~\ref{fig:CS-ALL-Region} represents the Consensus Ranking over all 10 HHS regions, based on the average of  consensus rankings across all \ERquantitiesb for each region listed in Table~\ref{tb:CS_all_regions} . As can be seen in Figure~\ref{fig:CS-ALL-Region}, the performance of the first and the second methods are behind the other approaches and we can exclude them from the pool of selected algorithm. However, the other four methods show very competitive performance  and are considered the same according the total rankings. The sequential aggregations provide a general conclusion which eliminates the nuances of similar methods.
}

\npdecimalsign{.}
\nprounddigits{1}
\begin{table*}[tb]
\centering
\captionsetup{font=scriptsize}
\caption{\label{tb:CS_all}  Average Consensus Ranking over different error \metricsb for all \ERquantities- \Revision{Region 1}. }
\npdecimalsign{.}
\nprounddigits{2}
\begin{tabular}{cp{0.8cm}p{0.8cm}p{0.8cm}p{0.8cm}p{0.8cm}p{0.8cm}p{1cm}p{1cm}p{0.8cm}p{0.8cm}}
\hline
  & {Peak value} & {Peak time} & {take-off-value}  & {take-off-time} & {ID length} & {ID start time} & {Start of flu season} & {Speed of Epidemic} & {Average} & {Median}\\
\hline 
M1 & 5.83 &	3.83 & 6 & 1 & 3.33	&  5.67 & 6 & 5.83 & 4.69 & 5.67	\\
\hline
M2 & 4.17  & 4.5 &  5 & 2	& 1	&  4.33 & 5.0 & 4.5  & 3.81 & 4.33\\
\hline
M3 & 3 &	2.83 & 3.83 & 3 & 3.33	&  3.17 &  3 & 3.17	& 3.17 & 3.17\\
\hline
M4 & 1.17 &	3.33 & 1.17 & 5 & 4.00	&  1.0 &  1 & 1.17	& 2.23 & 1.17\\
\hline
M5 & 4.17 &	1.17 & 3 & 4 & 4.33	&  4.67 &  3 & 4.17 & 3.56 & 4\\
\hline
M6 & 2.17 &	2.33 & 1.50 & 6 & 4.67	&  2.00 &  1.00 & 1.67	& 2.67 & 2.17 \\
\hline

\end{tabular}

\end{table*}

\npdecimalsign{.}
\nprounddigits{1}
\begin{table*}[tb]
\centering
\captionsetup{font=scriptsize}
\caption{\label{tb:CS_all_regions} \Revision{ Average Consensus Ranking of methods over different \ERquantities- Regions 1 - 10}. }
\npdecimalsign{.}
\nprounddigits{2}
\begin{tabular}{cp{0.8cm}p{0.8cm}p{0.8cm}p{0.8cm}p{0.8cm}p{0.8cm}p{1cm}p{1cm}p{0.8cm}p{0.8cm}p{0.8cm}}
\hline
  & {Region1} & {Region2} & {Region3}  & {Region4} & {Region5} & {Region6} & {Region7} & {Region8} & {Region9} & {Region10} &{Ave}\\
\hline 
M1 & 4.69 &	3.31 & 4.6 & 3.94 & 3.65&  2.21 & 4.3  & 3.94 & 3.46 & 4.29 & 3.84	\\
\hline
M2 & 3.81 & 2.77 & 4.23 & 4.0	& 3.71&  1.29 & 3.73 & 3.69  & 3.79 & 3.96 & 3.50\\
\hline
M3 & 3.17 &	3.46 & 1.96 & 2.68 &2.67&  2.21 &  3.03 & 2.73 & 2.17 & 2.33 & 2.64\\
\hline
M4 & 2.23 &	3.19 & 2.04 & 2.7  &3.08&  1.29 &  2.93 & 2.60 & 2.44 & 3.71 &2.62\\
\hline
M5 & 3.56 &	1.79 & 1.79 & 2.41 &2.77&  2.21 &  2.67 & 3.06 & 2.88 & 2.67 &2.58 \\
\hline
M6 & 2.67 &	3.23 & 2.13 & 2.48 &2.83&  1.29 &  2.60 & 3.27	& 3.13 & 3.58 & 2.72\\
\hline

\end{tabular}

\end{table*}

\subsection*{Horizon Rankings for each \ERquantity}
Horizon Ranking helps track the change in accuracy and ranking 
of the methods over prediction time. If the Horizon Ranking fluctuates a lot 
over the time steps, this hints at the unsuitability of Consensus Ranking as 
selection criteria for the best method. It is possible that the method that 
performs best during early stages of prediction need not perform the best at a 
later time-points. Figure \ref{fig:HorizonPeakV} shows the evolution of  
Horizon Ranking of the six methods for predicting the peak value calculated 
based on AEP and sAPE. As shown in Figure \ref{fig:AllEpiBoxPlot}, Methods 4 
and 6 have the best average consensus ranking in predicting peak value and is 
consistent with observations on Horizon ranking. In 
Figure \ref{fig:HorizonPeakV} the ranking of Methods 4 and 6 demonstrates a 
little fluctuation at the first time-steps. However, as prediction time moves 
forward they provide more accurate forecasts causing them to rank higher. 
The most interesting case for Horizon Rankings concerns the prediction of peak 
time. The Consensus Ranking in Figure \ref{fig:AllEpiBoxPlot}  selects Method 5 
as the superior in predicting peak time and Methods 6 and 4 as the second and 
third best approaches. However, by observing the trends of ranks over 
prediction times (Figure \ref{fig:HorizonPeakT}), Methods 4 and 6 are the 
dominant for the first eight weeks of prediction, and then method 1 wins the 
first place for seven weeks. In the next eight weeks, methods 1, 3, and 5 are 
superiors simultaneously. 
\\Figure \ref{fig:HorizonID} ,\ref{fig:HorizonTakeoff}, \ref{fig:HorizonOthers} 
shows Horizon Ranking graphs for leveraging forecasting methods in predicting 
other \ERquantities. These Horizon rankings are almost consistent with their 
corresponding Consensus ranking which confirms the superior methods from 
Consensus ranking perspective could be used for any prediction time. 

\subsection*{Visual Comparison of forecasting methods}
In order to visualize the output of forecasting methods, we generate the one-step-ahead epidemic curve. It means given the early time series up to time $k$ ($y\left(1\right),..., y\left(k\right)$) as observed data, the forecasting algorithm predicts the
the next data point of time series $x\left(k+1\right)$ and this process is repeated for all values of prediction time $k$ where $  t_{b} \leq k \leq t_e $. By putting together the short-term predictions, we construct a timeseries from  $t_{b}$ to $  t_{e}$ as one-step-ahead predicted epidemic curve. Figure~\ref{VisualR1} depicts the one-step-ahead predicted epidemic-curve for HHS region 1 that are generated by the six forecasting methods (refer to figures S11-S19 in Supplementary Information for other Regions).  We have used $t_{b} = 2$ and $  t_{e} = T-1 $ as the beginning and end for the prediction time. As can be seen in figure~\ref{VisualR1}, the first and second methods show bigger deviations from observed curve especially in the first half of the season.
As these six methods are different configurations of one algorithm, their outputs are so competitive and sometimes similar to each other. Methods 3 and 5, and methods 4 and 6 show some similarity in their one-step-ahead epidemic curve that is consistent with Horizon Ranking charts for various \ERquantities. However, Horizon Ranking graphs contains more information regarding the long-term predictions; Therefore, the ranking methods, especially Horizon Ranking, could help the experts to distinguish the better methods 
 when the outputs of forecasting methods are competitive and judging based on the visual graph is not straightforward. 
%

\section*{Epidemic Forecast Evaluation Framework}
We have proposed a set of \ERquantitiesb and error \metricsb and showed how to evaluate different forecasting methods together. These are incorporated into Software Framework as described (Figure \ref{fig:framework}).  The software framework, named Epi-Evaluator, receives the observed epidemic curve and predicted ones as an input and can generate various rankings based on the choice of \ERquantitiesb and error \metric. The system is designed as a collection of scripts that are loosely coupled through the data they exchange. This is motivated by two possible scenarios: (a) individuals must be able to use each module in isolation. (b) Users must not be restricted to the possibilities described in this paper, and be able to contribute \quantitiesb and \metricsb of their interest. 

We also include a standardized visualization module capable of producing a variety of plots and charts summarizing the intermediate outputs of each module. This enables the package to have a \emph{plug-and-play} advantage for end users. We envision the end-users ranging from (a) epidemiologists who wish to quickly extract/plot key \ERquantitiesb from a given surveillance curve, (b) computational modelers who wish to quantify their predictions and possibly choose between different modeling approaches, (c) forecasting challenge organizers who wish to compare and rank the competing models, and (d) policymakers who wish to decide on models based on their \ERquantityb of interest.

\section*{Evaluating Stochastic forecasts} 
\label{sec:prob}
The aforementioned \metricsb deal primarily with deterministic forecasts, whereas there are a lot of stochastic forecasting algorithm with some levels of uncertainty that makes the evaluation much  harder. Moreover, the observed data may be stochastic because of possible errors in measurements and source of information. We are going to extend our \metricsb and provide new methods to handle the stochastic forecasts and observation. Non-deterministic forecasts could be in one of the following formats:

\begin{itemize}
\item{Multiple replicates of the time series}
\item{A timeseries of mean and variance of the predicted values}
\end{itemize}

\subsection*{Stochastic forecasts as multiple replicates}
Most of the stochastic algorithms, generate multiple replicates of series and  state vectors to simulate the posterior density function by aggregating 
discrete values together. State vector contains the parameters that are used by the 
epidemic model to generate the epidemic curve (timeseries of new infected 
cases). Therefore, the best state vectors (models) are those that generate 
epidemic-curve closer to observed one (i.e., models with the higher likelihood). 
When the forecasting methods output is a 
collection of replicates of state vectors and timeseries, we have the option to 
calculate \ERquantitiesb on each series, for each prediction time, and assess 
the error \metricsb on each series. The error \metricsb should be accumulated 
across the series through getting Arithmetic Mean, Median, Geometric Mean, etc. 
to provide a unique comparable value per each method. Table~\ref{tb:metricsND} 
provides  advanced error \metricsb to aggregating the simple EM values over the 
series.

\begin{table*}[h!]
\centering
\caption{\label{tb:metricsND}  List of advanced error \metricsb to aggregating the simple EM values across multiple series.  }
\begin{tabular}{p{3cm}|c|p{4cm}}
\hline 
\Metricb name & Formula & Description \\
\hline
Absolute Percentage Error ($APE_{t,s}$) & $ APE_{t,s}=|\frac{y_{t} - x_{t,s}}{y_{t}}|  $ & where $t$ is time horizon and $s$ is the series index.  \\

Mean Absolute Percentage Error ($MAPE_{t}$) & $ MAPE=\frac {1}{S} \sum_{s=1}^S APE_{t,s}  $ & where $t$ is time horizon, $s$ is the series index $S$ is the number of series for the method. \\
 
Median Absolute Percentage Error ($MdAPE_{t}$) & Median Observation of $MdAPE_{t}$ & where Observations are sorted 
$ APE_{t,s} $ s, $t$ is time horizon, $s$ is the series index. \\
\hline
Relative Absolute  Error ($RAE_{t,s}$) & $ RAE_{t,s}=\frac{|y_{t} - x_{t,s}|}{|y_{t} - x_{RW_{t,s}}|}  $ & Measures the average ratio of absolute error to Random walk error in time horizon t.  \\
Geometric Mean Relative Absolute  Error ($GMRAE_{t}$) & $ GMRAE_t= [\prod_{s=1}^{S} |RAE_{t,s}| ]^{1/S}  $ & Measures the average ratio of absolute error to Random walk error  \\

Median Relative Absolute  Error ($MdRAE_{t}$) &  Median Observation of $RAE_{s} $  & Measures the median observation of sorted $RAE_{s}$  for time horizon t \\
\hline
Cumulative Relative Error ($CumRAE_s$ ) & $ CumRAE_s =\frac{\sum_{t=1}^T |y_{t,s} - x_{t,s}|}{\sum_{t=1}^T|y_{t,s} - x_{RW_{t,s}}|} $ & Ratio of accumulation of errors to cumulative error of Random walk Method  \\

Geometric Mean Cumulative Relative Error ($GMCumRAE$ ) & $ GMCumRAE =[\prod_{s=1}^{S} |CumRAE_s| ]^{1/S}  $ & Geometric Mean of Cumulative Relative Error across all series.\\

Median Cumulative Relative Error ($MdCumRAE$ ) & $ MdCumRAE = Median(|CumRAE_s|)  $ & Median of Cumulative Relative Error across all series.\\
\hline 
Root Mean Squared Error ($RMSE_t$) & $ RMSE_t= \sqrt{\frac{\sum_{s=1}^S (y_{t} - x_{t,s})^2}{S} } $ & Root square of average squared error across series in time horizon t\\
\hline
Percent Better ($PB_t$)    & $ PB_t=\frac {1}{S} \sum_{s=1}^S [I\{e_{s,t},e_{WRt}\}] $  & Demonstrates average number of times that method overcomes the Random Walk method in time horizon t.  \\
& $ |e_{s,t}|\leq|e_{WRt}| \leftrightarrow I\{e_{s,t},e_{WRt}\}=1  $&\\
\hline

\end{tabular}

\end{table*}

Armstrong \cite{Armstrong1992}, performed an evaluation over some of these \metricsb and suggested the best ones in different conditions. In calibration problems, a sensitive error measure is needed to demonstrate the change in parameters in the error \metricb values. The EMs with good sensitivity are: RMSE, MAPE, and GMRAE. He suggests GMRAE because of poor reliability of RMSE  and he claimed that MAPE is biased towards the low forecasts \cite{Armstrong1992}. As we mention in the discussion section, We believe that MAPE is not biased in favor of the low forecasts and could also be a good metric for calibration (Refer to Discussion section). Also, GMRAE could drop to zero when only one zero in the errors pops up and lower down the sensitivity of GMRAE to zero. 

For selecting among forecasting methods, Armstrong offered MdRAE when the output has a small set of series and MdAPE for a moderate number of series. He believes that reliability, protection against outliers, construct validity, and the relationship to decision-making are more important criteria than sensitivity. MdRAE is reliable and has better protection against outliers. MdAPE has a closer relationship to decision making and is protected against outliers \cite{Armstrong1992}. 

For the stochastic algorithms that generate multiple series with uneven weights, it is important to consider the weight of the series in calculating the arithmetic means. As an illustration, instead of calculating MAPE, sMAPE, RMSE, and MdAPE across the series, we suggest measuring weighted-MAPE, weighted-sMAPE, weighted-RMSE, and weighted-MdAPE respectively. 

\subsection*{Stochastic forecasts with uncertainty estimates}
Sometimes the output of stochastic forecasting method is in the form of mean 
value and variance/uncertainty interval for the predicted value. 
In statistics theory, the summation of 
Euclidean distance between the data points and a fixed unknown point in 
n-dimensional space is minimized in the mean point. Therefore, the mean value 
is a good representative of other data points. As a result, we can simply 
calculate epi-measure on the predicted mean value of epidemic curve and compare 
them through error metrics. However, this comparison is not comprehensive 
enough because the deviation from the average value is not included in the 
discussion. To handle this kind of evaluation, we divide the problem to two sub-problems
\begin{itemize}
\item{A) Deterministic observation and stochastic forecasts with uncertainty estimates }
\item{B) Stochastic observation and stochastic forecasts with uncertainty estimates}
\end{itemize}

\subsection*{A) Deterministic observation and stochastic forecasts with uncertainty estimates} 
\Revision{
In this case, we assume that each forecasting method's output is a timeseries of uncertain estimates of predicted case counts and is reported by the mean value $\overline{x_t}$, variance $\sigma_t^2$ for data point at $t^{th}$ week, and the number of samples $N_x$. For simplicity, we eliminate the subscript $t$. Table~\ref{tb:notations2} lists the required notations used in the following sections. Sample size refers to the number of predicted samples from which the mean and variance are obtained. In the best situation, the forecast algorithm could provide with the probability density function (pdf) of each predicted data point denoted by $f(x)$, unless we assume the pdf is Normal distribution $f_x \sim
 N(\mu_x,\sigma_x)$ for the large enough sample size, or t-distribution $f_x \sim
 t(\mu_x,v)$  if the sample size is low. T-distribution has heavier tails, which means it is more subject to producing values far from the mean.  $N_x \geq 30$ is assumed as large sample size. $N_x$ is used to calculate the standard deviation of the random variable X, from the standard deviation of its samples: $\sigma_x={\sigma}/\sqrt{N_x}$. When the sample size is low, the degree of freedom of t-distribution is calculated by $N_x$: $v=N_x - 1$.   }

\begin{table*}[h!]
\centering
\caption{\label{tb:notations2} \Revision{ Notation Table II }  }
\begin{tabular}{cp{10cm}}
\hline 
\textbf{Symbol} &   \textbf{Definition}
 \\
\hline
$X$  & Random variable $X$ (or $X_t$) that is the predicted estimate of a data point at one week( $t^{th}$ week)  \\
\hline
$f(x) | f_x$  & Probability density function (pdf) of random variable $X$  \\
\hline
$\mu_x$  & Mean value for the random variable $X$  \\
\hline
$\sigma_x = \sigma /{\sqrt{N_x}}$  & Standard deviation for the random variable $X$  \\
\hline
$\overline{x}$ & Mean value of the samples belonging to random variable $X$  \\
\hline
$\sigma$  & Standard deviation of the samples belonging to random variable $X$ \\
\hline
$v $  & $v=N_x - 1 $ Degree of freedom of t-distribution \\
\hline
$ \bar{y} $    & $\bar{y}=\frac {1}{n} \sum_{t=1}^n (y_t) $ : the mean for \textit{y} values over n weeks 
\\
\hline
$S_x=\left\{s_i\right\}$ & where $ s_i $ is the sample from distribution $f_x$\\
\hline
$N_{s_x} = |S_x| $ & Number of sample set $S_x$  \\ 
\hline 
$Y$  & Random variable $Y$ (or $Y_t$) that is the estimate of observed value at one week( $t^{th}$ week)  \\
\hline
$g(y)|g_y  $ & Probability density function (pdf) of random variable $Y$ \\ 
\hline 
$S_y=\left\{s_j\right\} $ &  where $ s_j $ is the sample from distribution $g_x$\\
\hline\ 
\end{tabular}
\end{table*}

\Revision{
In order to evaluate the performance of stochastic methods, we suggest to perform the Bootstrap sampling from the distribution $f(x)$ and generate the sample set $S_x=\{s_i\} $ for each data point of timeseries where $|S_x|>>N_x $. Note that we don't have access to the instances of the first sample size, so we generate a large enough sample set from its pdf function $f(x)$. Then the six selected error \metricsb, MAE, RMSE, MAPE, sMAPE, MdAPE, and MdsAPE, are calculated across the sample set $S_x$ \textit{for each week}. Table S8 in supplementary information contains the extended formulation of the error \metricsb used for stochastic forecasts.
Using the equations in Table S8 we can estimate different expected/median errors for each week for a stochastic forecasting method. The weekly errors could be aggregated my getting Mean or Median across the time to calculate the total error \metricsb for each method. The aggregated error \metricsb can be used to calculate the Consensus Ranking for the existing forecasting approaches. Moreover, having the errors for each week, we can depict the Horizon-Ranking and evaluate the trend of rankings across the time similar to the graphs for deterministic approaches. }

\subsection*{B) Stochastic observation and stochastic forecasts with uncertainty estimates} 
 There are a lot of source of errors in measurements and data collections which result in the uncertainty for the observation data and makes the evaluation more challenging. We suggest two categories of solutions to deal with this problem:
\begin{itemize}
\item{a) Calculating the Distance between probability density functions}
\item{b) Calculating the error \metricsb between two probability density functions }
\end{itemize}

\subsection*{B-a) Calculating the Distance between probability density functions} 
\Revision{
Assuming that both predicted and observed data are stochastic, they are represented as the timeseries of probability densities functions (pdfs). There are a lot of distance functions that can calculate the distance between two pdfs \cite{R-Distance}. Three of the most common distance functions for this application are listed in Table~\ref{tb:pdfDistance}.  
}

\begin{table*}[h!]
\centering
\caption{\label{tb:pdfDistance}  \Revision{ Distance functions to measure dissimilarity between probability density functions of stochastic observation and stochastic predicted outputs.} }
\begin{tabular}{p{1.7cm}|c|c}
\hline 
Distance Function & Formula (Continues) & Formula (Discrete form)  \\
\hline
Bhattacharyya  & $$$ D_B(P,Q)=-Ln(BC(P,Q)) $$$ &  $$$ D_B(P,Q)=-Ln(BC(P,Q)) $$$  \\
&, $$$BC(P,Q)=\int \sqrt{P(x)Q(x)}dx$$$& $$$ ,BC(P,Q)=\sum \sqrt{P(x)Q(x)}$$$   \\
\hline
Hellinger & $$$ D_H=\sqrt { 2\int {(P(x)-Q(x))^2}dx} $$$ &  $$$ D_H(P,Q)=\sqrt { 2\sum_{k=1}^{d} {(P(x_k)-Q(x_k))^2}}  $$$    \\
& $$$= 2\sqrt{1-\int \sqrt{P(x)Q(x)}dx}$$$& $$$ = 2\sqrt{1-\sum_{k=1}^{d} \sqrt{P(x_k)Q(x_k)}}$$$   \\
\hline 
Jaccard &  - &  $$$ D_{Jac} = 1- S_{Jac} $$$   \\
&  & $$$ S_{Jac}  = \frac{\sum_{k=1}^{d}{P(x_k)\times Q(x_k)}}{ \sum_{k=1}^{d}{P(x_k)^2} + \sum_{k=1}^{d}{Q(x_k)^2}- \sum_{k=1}^{d}{P(x_k).Q(x_k)}}$$$   \\
\hline
\end{tabular}
\end{table*}

Bhattacharyya distance function \cite{R-Distance} and Hellinger \cite{Hellinger} both belong to the squared-chord family, and their continues forms are available for comparing continues probability density functions. In special cases, like when the two classes are under the normal distribution, these two distance functions can be calculated by the mean and variances of pdfs as follows \cite{Normal_Bhat,NormalHellinger}:

\begin{eqnarray}
\label{eq:Bha}
\centering
D_{B}(P,Q) = \frac{1}{4}ln \left(\frac{1}{4} \left(\frac{\sigma_p^2}{\sigma_q^2} +\frac{\sigma_q^2}{\sigma_p^2}+2 \right) \right) + \frac{1}{4}\left(\frac{(\mu_p-\mu_q)^2}{\sigma_p^2 + \sigma_q^2} \right)  
\end{eqnarray}

\begin{eqnarray}
\label{eq:Hellinger}
\centering
D_{H}^2(P,Q) = 2 \left(1- \sqrt{\frac{2\sigma_1.\sigma_2 }{\sigma^2_1+\sigma_2^2}}.exp(\frac{-(\mu_1-\mu_2)^2}{4(\sigma^2_1+\sigma_2^2)}) \right) 
\end{eqnarray}

However, calculating the Integral may not be straightforward for every kind of pdfs. Also, Jaccard distance function is in the discrete form. To solve this problem, we suggest Bootstrap sampling from both predicted and observed pdfs and generating the sample set $S = S_x \cup S_y $  where $S_x=\left\{s^x_i|s^x_i\sim f(x)\right\}$, $S_y=\left\{s_j^y|s_j^y\sim g(y)\right\}$, and  $|S_x|=|S_y|>>N_x $. Then we calculated the summation for the distance function over all the items that belong to the sample set $S$. As an example for Jaccard distance function:

\begin{eqnarray}
\label{eq:Jaccard}
\centering
D_{Jac} = 1-\frac{\sum_{k=1}^{|S|}{f(s_k)\times g(s_k)}}{ \sum_{k=1}^{|S|}{f(s_k)^2} + \sum_{k=1}^{|S|}{g(s_k)^2}- \sum_{k=1}^{|S|}{f(s_k)\times g(s_k)}}
\end{eqnarray}

\Revision{
Jaccard distance function belongs to inner product class and incorporates both similarity and dissimilarity of two pdfs. Using one of the aforementioned distance functions between the stochastic forecasts and stochastic observation, we can demonstrate Horizon Ranking across time and also aggregate the distance values by getting the mean value over the weeks and then calculate the Consensus Ranking. 
\\Although these distance functions between the two pdfs seem to be a reasonable metric for comparing the forecast outputs, it ignores some information about the magnitude of error and its ratio to the real value. In other word, any pair of distributions like (P1,Q1) and (P2,Q2) could have same distance value if : $|\mu_{P_1}-\mu_{Q_1}| = |\mu_{P_2}-\mu_{Q_2}|$ and $\sigma_{P_1}=\sigma_{P_2}$ and $\sigma_{Q_1}=\sigma_{Q_2}$. Therefore, the distance functions do not consider the domain values of ($x_i$, $y_j$) and lose the information about the relative magnitude of error to the observed value. 
 In the ranking process of different forecasting approaches, as the observed data is assumed to be fixed, this issue will not be a concern. 
The other problem of using distance functions between pdfs arises when some forecasting methods are stochastic, and others are deterministic. As the proposed error \metricsb are not compatible with distance functions, we cannot compare them together.
 }

\subsection*{B-b) Calculating the error \metricsb between two probability density functions }
\Revision{
In order to compare stochastic and deterministic forecasting approaches together, we suggest estimating the same error \metricsb used for deterministic methods. We perform Bootstrap sampling from both predicted and observed pdfs for each data point of time-series and generate two separate sample sets $S_x$ and $S_y $ where $S_x=\left\{s^x_i|s^x_i\sim f(x)\right\}$, $S_y=\left\{s_j^y|s_j^y\sim g(y)\right\}$ and $|S_x|=|S_y|>>N_x $. The  six selected error \metrics, MAE, RMSE, MAPE, sMAPE, MdAPE, and MdsAPE , could be estimated through the equations listed in Table S9 in supporting information. These \metricsb incorporate the variance of pdfs through the sampling and represent the difference between the predicted and observed densities by weighted expected value of the error across the samples. 
}

\section*{Discussion}
As shown in previous sections, none of the forecasting algorithms may 
outperform the others in predicting all \ERquantities. For a given \ERquantity, 
we recommend using the consensus ranking across different error \metrics. 
Further, even for a single \ERquantity, the rankings of methods seem to vary as 
the prediction time varies. 

\subsection*{Horizon Ranking vs Consensus Ranking}
How do we decide on the best method when Horizon ranking and Consensus ranking 
lead to different conclusions? The significant difference between Horizon and 
Consensus Rankings comes from the fact that Consensus Ranking calculates the 
average (or median) of the errors for a given time step and then sorts them to 
determine the ranking. This aggregation of errors is not always a disadvantage,
because sometimes a slight difference in errors could change
the Horizon Ranking level while the Consensus Ranking accumulates the errors 
for whole time-series which gives an overall perspective of methods' 
performance. If the purpose of evaluation is to select a method as the best 
predictor for all weeks, Consensus rankings can be used to guide the method 
selection. However, if there is a possibility for using different prediction 
methods at different periods, we offer to determine a few time 
intervals in which the Horizon Rankings of the best methods are consistent. 
Then, in each time interval, the best method based on Horizon Ranking could be 
selected, or the Consensus Ranking could be calculated for each period by 
calculating the average errors (error \metrics) over time steps. The 
superior method for each time interval is the one with first Consensus Ranking in 
that period.  One of the advantages of Horizon Ranking is to detect and reduce 
the effect of outliers across time horizons, whereas Consensus Ranking 
aggregates the errors across time steps that results in a noticeable change in 
total value of error \metricsb by outliers.  

\subsection*{MAPE vs sMAPE}
MAPE and sMAPE have been the two important error \metricsb for measuring forecast 
errors since 1993. MAPE was used as the primary measure in M2-Competition, and 
it was replaced by sMAPE in M3-Competition to overcome the disadvantages of 
MAPE. One of the drawbacks is that MAPE could get a large or undefined value 
when the observed data point gets close to zero. That's one of the reasons why sMAPE used the average of observed and predicted value in the denominator 
to alleviate this phenomenon. The other issue that has been claimed for MAPE in 
some literature is biasing in favor of small forecasts. Therefore, the critics 
believe that MAPE leads to higher penalty for
large overestimation rather than any underestimation. sMAPE, as the symmetric version of 
MAPE, normalized the error value with the mean of predicted and observed data 
which limits the range of sMAPE error between 0 and 2 for both overestimating 
and underestimating of the prediction. 
However, we believe that although the range of sMAPE function is symmetric, it 
does not provide a uniform scoring of the errors. We believe sMAPE is 
significantly biased toward the large forecasts. 
Figure~\ref{fig:VS1} and Table S8 in supporting 
information demonstrate the corresponding domains that generate equal MAPE or 
sMAPE errors in term of magnitude. The figures in the left column belong to MAPE 
and the right ones are sMAPE's. In figure~\ref{fig:VS1}, the black line 
represents the observed epidemic curve (y), and the horizontal axis is the 
weekly time steps (t). The yellow borders show the predicted 
curves as overestimated 
or underestimated predictions which both results in MAPE= 0.5 or sMAPE = 0.5. The green spectrum shows the 
predicted curves with low values of MAPE or sMAPE. Equal colors in these figures correspond to equal values for the discussed error \metric. The red borders in the left graph belong to predicted curves $x(t)=2 \times y(t)$ and $x(t)=0 \times y(t)$ with MAPE = 1 and the red borders in the right chart corresponds to $x(t)=3\times y(t)$ and $x(t)=(1/3)\times y(t)$ which generate sMAPE = 1. As can be seen, MAPE grows faster than sMAPE which means MAPE reaches to 1 with smaller values in the domain. Moreover, MAPE demonstrates symmetrical growth around the observed curve that results in fair scoring toward over and underestimation. 

The black borders in lower charts are corresponding to predicted 
epidemic curve which generates MAPE=2 and sMAPE =2 in the left and right charts 
sequentially.  The color spectrum of sMAPE in the right chart represents the 
non-symmetric feature of this error \metricb which is in favor of large 
predictions. As we couldn't show the infinity domain for sMAPE, we limited 
it to the predicted curve $x(t)=20 \times y(t)$. Figure~\ref{fig:VS4} shows the 
blue spectrum of MAPE that corresponds to large predictions where $x(t)>>3y(t)$ 
and MAPE approaches infinity. This error \metricb provides more sensible 
scoring for both calibration and selection problems.

\subsection*{Relative evaluation vs Absolute one }
In this paper, we covered how to evaluate the performance of forecasting algorithm relative to each other and rank them based on various error measures. The ranking methods, like the Horizon ranking, can represent the difference in performances, even when the algorithms are so competitive. However, the other question that arises is, how can we recognize that the ranked methods have similar performance or that they are completely far from each other? What if we only have one forecasting method and just need to know about its performance? 
\\ The absolute measurement is a bigger challenge because most of the available error \metricsb are not scaled or normalized and do not provide meaningful range. However, if you have only one forecasting method to evaluate, we suggest taking advantage of MAPE \metric, as it is scaled based on the observed value and its magnitude defines how large on average the error is, compared with the observed value. 
\\For multiple algorithms, we suggest to calculate MAPE \metricb on the one-step-ahead epidemic curve of each algorithm and cluster them based on its MAPE value. As discussed in the previous section and Table S10, four meaningful intervals for MAPE value could be defined as the criteria to cluster the forecasting approaches into the four corresponding groups which means: Methods with  $ 0 \leq MAPE \leq 1/2$, Methods with  $ 1/2 \leq MAPE \leq 1$, Methods with  $ 1 \leq MAPE \leq 2$, and Methods with  $ 2 \leq MAPE $. This kind of clustering can provide borders between the methods which are completely different in the performance. Then the algorithms of each group can be passed through the three steps of evaluation framework, and be ranked based on various \ERquantitiesb and error \metrics. As an illustration, Table~\ref{tb:MAPEcluster} provides the average value of different error \metricsb over all 10 HHS regions for the six aforementioned methods and an autoregressive forecasting method named ARIMA \cite{ref:arima}. As can be seen, the MAPE value of the six methods are under 0.5, that clusters all of them in the same category, while the MAPE for the ARIMA method is 0.77 which assigns it to the second group; It means the performance of ARIMA is completely behind all other methods. Figure~\ref{Arima} depicts the 1-step-ahead predicted curve of ARIMA method comparing to the observed data that shows ARIMA output has large deviations from the real observed curve and confirms the correctness of clustering-approaches. 

\begin{table*}[tb]
\centering
\captionsetup{font=scriptsize}
\caption{\label{tb:MAPEcluster} \Revision{ Different error \metricsb calculated for one-step-ahead epidemic curve over whole season (2013-2014), averaged across all HHS regions: Comparing Methods M1 to M6 and ARIMA approach.} }
\begin{tabular}{ccccccc}
\hline
         & {MAE } & {RMSE} & {MAPE }  & {sMAPE } & {MdAPE} & {MdsAPE} \\
\hline
Method 1 &  316.18 & 378.63 & \textbf{0.39}  & 0.33 & 0.34 &  0.29  \\
\hline
Method 2 & 293.76 & 357.34 & \textbf{0.35}  & 0.31 & 0.30  &  0.26 \\
\hline
Method 3 & 224.53 & 293.52 & \textbf{0.25}  & 0.22 & 0.22 &  0.20 \\
\hline
Method 4 & 204.5 &  274.41 & \textbf{0.21}  & 0.21 & 0.18 &  0.18 \\
\hline
Method 5 & 224.57 & 293.90 & \textbf{0.25 } & 0.22	& 0.22 & 0.20 \\
\hline
Method 6 & 204.25 &	274.97 & \textbf{0.21 }&	0.20 &	0.18 &	0.18 \\
\hline
ARIMA & 1015.60 & 1187.62& \textbf{0.77}  & 0.74 & 0.78 & 0.75 \\
\hline

\end{tabular}

\end{table*}

\subsection*{Prediction Error vs Calibration Error}
In this paper, prediction error is considered to calculate the \textit{predicted} error measures, i.e. only the errors after prediction time is taken into account and the deviation between the model curve and data before prediction time is ignored. However, we suggest the evaluator framework in two different modes: Forecasting mode vs Calibration mode. As mentioned in the forecasting mode, only prediction error is measured. Moreover, if the observed \ERquantityb is already occurred in the $i^{th}$ week, the forecasts corresponding to the prediction times after the $i^{th}$ week are not considered in accumulation of the errors because they are not interested anymore. However, in calibration mode, the aim is to find the error between model curves and observed data, regardless of the time of observed \ERquantity. Therefore the error measures on one epi-feature are accumulated for all prediction weeks.  Also, in calculating error \metricsb on the epidemic curve, the fitting errors before the prediction time are cumulated with prediction errors, to measure the calibration one.

\section*{Conclusion}
Evaluating epidemic forecasts arising from varied models is inherently 
challenging due to the wide variety of epidemic \quantitiesb and error 
\metricsb to choose from. We proposed different \ERquantitiesb for 
quantifying the prediction accuracy of forecasting methods and demonstrated how
suitable error \metricsb could be applied to those \ERquantitiesb to evaluate
the accuracy and error of prediction. We have applied the proposed \ERquantitiesb 
and error \metricsb on the output of six forecasting methods to assess their 
performance. As the experimental results showed, different error \metricsb 
provides various measurements of the error for a single \ERquantity. 
Therefore, we provided the Consensus ranking method to aggregate the rankings 
across error \metricsb  and summarize the performance of forecasting algorithms 
in predicting a single \ERquantity. Based on the first round of rankings, none 
of the forecasting algorithms could outperform the others in predicting  all 
\ERquantities. Therefore, we recommended the second set of rankings to 
accumulate the analysis for various \ERquantitiesb and provide a total summary 
of forecasting methods' capabilities.  We also proposed Horizon ranking to 
trace the performance of algorithms across the time steps to provide better 
perspective over time. We finally hint at how these methods can be adapted for 
the stochastic setting. 
Choosing the best forecasting method enables policy planners to make more 
reliable recommendations. Understanding the practical relevance of various 
\ERquantitiesb of interest, and the properties offered by different error \metricsb 
will help guide the method selection. We hope that our work allows for a more 
informed conversation and decision process while using and evaluating epidemic 
forecasts.

\section*{Declarations}

\begin{backmatter}

\section*{List of abbreviations}
List of abbreviations are included in Table~\ref{tb:abbreviations}.
\begin{table*}[h!]
\caption{\label{tb:abbreviations}  List of Abbreviations.  }
\begin{tabular}{ll}
\hline 
Age-AR	& Age-specific Attack Rate \\
APE&	Absolute Percentage Error		\\
ARI&	acute-respiratory-illness 	\\
CDC	&Centers for Disease Control and Prevention\\
cMAPE&	corrected MAPE	\\
CR	&Consensus Ranking	\\
CumRAE&	Cumulative Relative Error	\\
DARPA&	Defense Advanced Research Projects Agency	\\
Epi-features&	epidemic features		\\
GM	&Geometric Mean\\
GMRAE&	Geometric Mean Relative Absolute Error\\
HHS&	Department of Health and Human Services \\
ID	&Intensity Duration		\\
ILINet	&Influenza-like Illness Surveillance Network	\\
MAAPE	&Mean Arctangent Absolute Percentage Error	\\
MAE&	Mean Absolute Error		\\
MAPE&	Mean Absolute Percentage Error \\
MARE&	Mean Absolute Relative Error 		\\
MASE&	Mean Absolute Scaled Error		\\
M-Competitions&	Makridakis Competitions	\\
Md&	Median	\\
MdRAE&	Median Relative Absolute Error	\\
NIH&	National Institutes of Health\\
NMSE&	Mean Normalized Mean Squared Error	\\
NOAA	&National Oceanic and Atmospheric Administration	\\
PB&	Percent Better 		\\
pdf	&probability density function	\\
RMAE	&Relative Measures	\\
RMSE	&root-mean-square-error		\\
sAPE&	symmetric Absolute Percentage Error\\
SAR	&Secondary Attack Rate		\\
sMAPE	&symmetric Mean Absolute Percentage Error 	\\
SpE	&Speed of Epidemic 		\\
TAR	&Total Attack Rate		\\
\hline
\end{tabular}

\end{table*}

\section*{Ethics approval and consent to participate}
Not applicable

\section*{Consent for publication}
Not applicable

\section*{Availability of data and material}
Data that was used during the study are available upon request.

\section*{Competing interests}
  The authors declare that they have no competing interests.

\section*{Funding}
This work has been supported by the Intelligence Advanced Research Projects Activity (IARPA) via Department of Interior National Business Center (DoI-NBC) contract number D12PC00337.
\section*{Author's contributions}
    FST and MM and NR has contribution in the main idea and the framework of the project. 
		FST proposed and implemented the \ERquantitiesb and Error \Metricsb and generated the outputs and Diagrams and interpreted them.
		JC and BL have analyzed and prepared the data for the Digital Library for the simulation. 
		FST wrote the paper. SV and PC were  major contributors in writing the manuscript. 
		All authors read and approved the final manuscript.


\section*{Acknowledgements}
  We thank DTRA CNIMS, IARPA, NetSE, NIH MIDAS and DTRA V\&V for their support and also thank our external collaborators and members of the Network Dynamics and Simulation Science Laboratory (NDSSL) for their suggestions and comments.
 \\ We thank Fereshteh Kazemi and Milad Hosseinipour for their invaluable support.


\bibliographystyle{bmc-mathphys} 
\bibliography{bmc_article-1V}      




\section*{Figures}
\begin{figure}[h!]
\caption{\csentence{Software Framework }: 
Software Framework contains four packages: \ERquantitiesb package, Error \Metricb 
Packages, Ranking schema and Visualization Module. The packages are independent and are only connected through the exchanged data. 
}
\label{fig:framework}
\end{figure}

\begin{figure}[h!]
	\caption{\csentence{Predicting Epidemic Curve.} The red arrow points to the prediction time $k$ in which prediction occurs based on $k$ initial data points of time-series. The red dashed line is predicted epidemic curve and the black line is observed one. }\label{fig:fig0}
\end{figure}

\begin{figure}[h!]
	\caption{\csentence{Figure explaining Intensity Duration} Intensity Duration's length (ID) indicates the number of weeks where the number of new infected case counts are more than a specific threshold.}\label{fig:fig1}
\end{figure}

\begin{figure}[h!]
\caption{\csentence{Figure explaining Speed of Epidemic} Speed of Epidemic (SpE) is
the steepness of the line that connects the start data-point of time-series sequence
to the peak data-point. SpE indicates how fast the infected case counts reach the
peak value.}\label{fig:fig2}
\end{figure}


\begin{figure}[h!]
\caption{\csentence{HHS region Map,} based on ``U.S. Department of Health \& Human Services''
division\cite{HHS-regions} }\label{fig:Map}
\end{figure}

\begin{figure}[h!]
\caption{\csentence{Box-Whisker Plot shows the Consensus Ranking} of forecasting methods in predicting \textbf{Peak value} for Region 1, aggregated on different error \metrics}\label{fig:peakBoxplot}
\end{figure}

\begin{figure}[h!]
\caption{\csentence{Consensus Ranking of forecasting methods over all error \metricsb for predicting different \ERquantitiesb for Region 1} Method 4 is superior in 
predicting five \ERquantitiesb out of eight ones,but is far behind other methods in predicting three other \ERquantities. } 
\label{fig:AllEpiBoxPlot}
\end{figure}

\begin{figure}[h!]
\caption{\csentence{The box-whisker diagrams} shows the median, mean and the variance of consensus ranking of methods over all \ERquantitiesb for Region 1.  }
\label{fig:CS_BoxPlot}
\end{figure}

\begin{figure}[h!]
\caption{\Revision{\csentence{Consensus Ranking over all Epi-Features - Regions 1-6.}
 The box-whisker diagrams show the median, mean and the variance of consensus ranking of methods in predicting different \ERquantities.  }}
\label{fig:CS-ALL-R1-6}
\end{figure}

\begin{figure}[h!]
\caption{\Revision{
\csentence{Consensus Ranking over all Epi-Features- Regions 7-10} The box-whisker diagrams show the median, mean and the variance of consensus ranking of methods in predicting different \ERquantities.}}
\label{fig:CS-ALL-R7-10}
\end{figure}

\begin{figure}[h!]
\caption{\Revision{\csentence{Consensus Ranking over all 10 HHS-Regions} The box-whisker diagrams show the median, mean and the variance of consensus ranking of methods in predicting the \ERquantitiesb for all HHS regions.}}
\label{fig:CS-ALL-Region}
\end{figure}

\begin{figure}[h!]
\caption{\csentence{Horizon Ranking} of six methods for predicting the peak value calculated based on AEP, and sAPE, on Region 1. } 
\label{fig:HorizonPeakV}
\end{figure}

\begin{figure}[h!]
\caption{\csentence{Horizon Ranking} of six methods for predicting the peak time calculated based on AEP, and sAPE, on Region 1. Methods 4 and 6 are the dominant for the first eight weeks of prediction, and then method 1 wins the first place for seven weeks. In the next eight weeks, methods 1, 3, and 5 are superiors simultaneously. 
 } 
\label{fig:HorizonPeakT}
\end{figure}

\begin{figure}[h!]
\caption{\csentence{Horizon Ranking} of six methods for predicting the Intensity Duration length and start time calculated based on AEP, and sAPE, on Region 1. } 
\label{fig:HorizonID}
\end{figure}

\begin{figure}[h!]
\caption{\csentence{Horizon Ranking} of six methods for predicting the Take-off value and time calculated based on AEP, and sAPE, on Region 1. } 
\label{fig:HorizonTakeoff}
\end{figure}

\begin{figure}[h!]
\caption{\csentence{Horizon Ranking graphs} for leveraging forecasting methods in predicting Speed of Epidemic and Start of flu season, on Region 1.
 } 
\label{fig:HorizonOthers}
\end{figure}

\begin{figure}[h!]
\caption{\csentence{\Revision{ Visual comparison of 1-step-ahead predicted curves generated by six methods vs. the observed curve, Region 1 } } \Revision{: 
  The first and second methods show bigger deviations from observed curve especially in the first half of the season.
	As the six methods are different configurations of one algorithm, their outputs are so competitive and sometimes similar to each other; methods 3 and 5, and methods 4 and 6 show some similarity in their one-step-ahead epidemic curve that is consistent with Horizon Ranking charts for various \ERquantities.
 } }
\label{VisualR1}
\end{figure}


\begin{figure}[h!]
\caption{\csentence{Comparison of MAPE and sMAPE domains and ranges spectrum  }: 
Red borders in the left graph (A) belong to predicted curves $x(t)=2 \times y(t)$ and $x(t)=0 \times y(t)$ with MAPE = 1 and the red borders in the right chart (B) corresponds to $x(t)=3\times y(t)$ and $x(t)=(1/3)\times y(t)$ which generate sMAPE = 1. The black borders in graphs C \& D are corresponding to predicted epidemic curves which generates MAPE=2 and sMAPE =2 in the left and right charts sequentially. 
}
\label{fig:VS1}
\end{figure}

\begin{figure}[h!]
\caption{\csentence{Colored Spectrum of MAPE range }: 
MAPE does not have any limitation from upper side that results in eliminating the large overestimated forecasting.
}
\label{fig:VS4}
\end{figure}

\begin{figure}[h!]
\caption{\csentence{\Revision{ 1-step-ahead predicted curve generated by ARIMA vs the observed curve } } \Revision{: 
  The large gap between predicted and observed curves shows that ARIMA performance is behind the other six approaches and confirms that clustering approach based on MAPE value could be a good criteria for discriminating methods with totally different performances.
 } }
\label{Arima}
\end{figure}



\section*{Tables}


\section*{Additional Files}
  \subsection*{Additional file 1 --- Supporting Information.pdf}
  This is a pdf file in which our forecasting algorithm and the six used configurations are described. 

  \subsection*{Additional file 2 --- Supporting\_Tables.pdf}
    This is a pdf file which contains 8 tables in support of the Figures \ref{fig:AllEpiBoxPlot} , \ref{fig:VS1} and \ref{fig:VS4}. 

  \subsection*{Additional file 3 --- S1 Fig.png}
    Summary of Methodology: This figure is referred in Supporting Information.pdf, describing the forecasting pipeline.  
	
		\subsection*{Additional file 4 --- S2 Fig.pdf}
    Consensus Ranking of forecasting methods over all error \metrics for predicting different \ERquantitiesb for Region 2 
		
		\subsection*{Additional file 5 --- S3 Fig.pdf}
    Consensus Ranking of forecasting methods over all error \metrics for predicting different \ERquantitiesb for Region 3

		\subsection*{Additional file 6 --- S4 Fig.pdf}
    Consensus Ranking of forecasting methods over all error \metrics for predicting different \ERquantitiesb for Region 4
	
		\subsection*{Additional file 7 --- S5 Fig.pdf}
    Consensus Ranking of forecasting methods over all error \metrics for predicting different \ERquantitiesb for Region 5
		
\subsection*{Additional file 8 --- S6 Fig.pdf}
    Consensus Ranking of forecasting methods over all error \metrics for predicting different \ERquantitiesb for Region 6
		
		\subsection*{Additional file 9 --- 7 Fig.pdf}
    Consensus Ranking of forecasting methods over all error \metrics for predicting different \ERquantitiesb for Region 7
		
		\subsection*{Additional file 10 --- S8 Fig.pdf}
    Consensus Ranking of forecasting methods over all error \metrics for predicting different \ERquantitiesb for Region 8
		
		\subsection*{Additional file 11 --- S9 Fig.pdf}
    Consensus Ranking of forecasting methods over all error \metrics for predicting different \ERquantitiesb for Region 9
		
		\subsection*{Additional file 12 --- S10 Fig.pdf}
    Consensus Ranking of forecasting methods over all error \metrics for predicting different \ERquantitiesb for Region 10
		%
		\subsection*{Additional file 13 --- S11 Fig.pdf}
    \Revision{ Visual comparison of 1-step-ahead predicted curves generated by six methods vs. the observed curve, Region 2. }
		
		\subsection*{Additional file 14 --- S12 Fig.pdf}
    \Revision{ Visual comparison of 1-step-ahead predicted curves generated by six methods vs. the observed curve, Region 3. }

		\subsection*{Additional file 15 --- S13 Fig.pdf}
    \Revision{ Visual comparison of 1-step-ahead predicted curves generated by six methods vs. the observed curve, Region 4. }
	
		\subsection*{Additional file 16 --- S14 Fig.pdf}
    \Revision{ Visual comparison of 1-step-ahead predicted curves generated by six methods vs. the observed curve, Region 5. }
		
\subsection*{Additional file 17 --- S15 Fig.pdf}
    \Revision{ Visual comparison of 1-step-ahead predicted curves generated by six methods vs. the observed curve, Region 6. }
		
		\subsection*{Additional file 18 --- S16 Fig.pdf}
    \Revision{ Visual comparison of 1-step-ahead predicted curves generated by six methods vs. the observed curve, Region 7. }
		
		\subsection*{Additional file 19 --- S17 Fig.pdf}
    \Revision{ Visual comparison of 1-step-ahead predicted curves generated by six methods vs. the observed curve, Region 8. }
		
		\subsection*{Additional file 20 --- S18 Fig.pdf}
    \Revision{ Visual comparison of 1-step-ahead predicted curves generated by six methods vs. the observed curve, Region 9. }
		
		\subsection*{Additional file 21 --- S19 Fig.pdf}
    \Revision{ Visual comparison of 1-step-ahead predicted curves generated by six methods vs. the observed curve, Region 10. }

\end{backmatter}
\end{document}